\pdfoutput=1
\documentclass{article}
\usepackage{enumerate}
\usepackage{graphicx}
\usepackage{hyperref}
\usepackage{lineno}
\usepackage[OT1]{fontenc}
\usepackage{amsthm,amsmath,amssymb,mathrsfs}
\usepackage[numbers]{natbib}
\usepackage{graphicx}
\usepackage{subfig}
\usepackage{titlesec}
\usepackage{times}
\usepackage{amsfonts}
\RequirePackage{enumitem} 
\RequirePackage{mathtools}
\usepackage[doublespacing]{setspace}
\usepackage{geometry}
\makeatletter
\renewcommand\@biblabel[1]{} 
\renewenvironment{thebibliography}[1]
{\section*{\refname}%
	\@mkboth{\MakeUppercase\refname}{\MakeUppercase\refname}%
	\list{\@biblabel{\@arabic\c@enumiv}}%
	{\settowidth\labelwidth{\@biblabel{}}%
		\leftmargin\labelwidth
		\advance\leftmargin15pt
		\advance\leftmargin\labelsep
		\setlength\itemindent{-10pt}
		\@openbib@code
		\usecounter{enumiv}%
		\let\p@enumiv\@empty
		\renewcommand\theenumiv{\@arabic\c@enumiv}}%
	\sloppy
	\clubpenalty4000
	\@clubpenalty \clubpenalty
	\widowpenalty4000%
	\sfcode`\.\@m}
{\def\@noitemerr
	{\@latex@warning{Empty `thebibliography' environment}}%
	\endlist}
\renewcommand\newblock{\hskip .11em\@plus.33em\@minus.07em}
\makeatother

\newenvironment{bottompar}{\par\vspace*{\fill}}{\clearpage}
\renewcommand{\refname}{REFERENCES}

\allowdisplaybreaks

\begin{document}

\title{\textbf{Blinded and unblinded sample size re-estimation procedures for stepped-wedge cluster randomized trials}}
\author{\textbf{M. J. Grayling\textsuperscript{1}, A. P. Mander\textsuperscript{1}, J. M. S. Wason\textsuperscript{1,2}}\\
\small 1. Hub for Trials Methodology Research, MRC Biostatistics Unit, Cambridge, UK\\ \small 2. Newcastle University, Newcastle, UK.}
\date{}
\maketitle

\noindent \textbf{Running Head:} Sample size re-estimation in stepped-wedge trials.\\

\noindent \textbf{Abstract:} The ability to accurately estimate the sample size required by a stepped-wedge (SW) cluster randomized trial (CRT) routinely depends upon the specification of several nuisance parameters. If these parameters are mis-specified, the trial could be over-powered, leading to increased cost, or under-powered, enhancing the likelihood of a false negative. We address this issue here for cross-sectional SW-CRTs, analyzed with a particular linear mixed model, by proposing methods for blinded and unblinded sample size re-estimation (SSRE). Blinded estimators for the variance parameters of a SW-CRT analyzed using the Hussey and Hughes model are derived. Then, procedures for blinded and unblinded SSRE after any time period in a SW-CRT are detailed. The performance of these procedures is then examined and contrasted using two example trial design scenarios. We find that if the two key variance parameters were under-specified by 50\%, the SSRE procedures were able to increase power over the conventional SW-CRT design by up to 29\%, resulting in an empirical power above the desired level. Moreover, the performance of the re-estimation procedures was relatively insensitive to the timing of the interim assessment. Thus, the considered SSRE procedures can bring substantial gains in power when the underlying variance parameters are mis-specified. Though there are practical issues to consider, the procedure's performance means researchers should consider incorporating SSRE in to future SW-CRTs. \\


\begin{bottompar}
	\noindent Address correspondence to M. J. Grayling, MRC Biostatistics Unit, Forvie Site, Robinson Way, Cambridge CB2 0SR, UK; Fax: +44-(0)1223-330365; E-mail: mjg211@cam.ac.uk. 
\end{bottompar}

\section{Introduction}

A stepped-wedge (SW) cluster randomised trial (CRT) involves the sequential roll-out of an intervention across several clusters over multiple time periods, with the time period in which a cluster begins receiving the intervention determined at random. Recent papers have established methods for sample size determination in the case of cross-sectional \citep{hussey2007} and cohort \citep{hooper2016} designs, for trials with multiple levels of clustering and for incomplete block SW-CRTs \citep{hemming2015}.

Undeniably, there has been a growing interest in the design, and in particular, it has now become associated with scenarios in which there is a belief that the trial's experimental intervention will be effective \citep{brown2006,mdege2011}. Given this commonly held belief, it may come as a surprise that a recent literature review determined that in 31\% of the SW-CRTs completed by February 2015, there was no significant effect of the experimental intervention on any of the trials primary outcome measures \citep{grayling2017}. To guard against this, implicitly assuming this failure rate was due to the experimental interventions being futile, methodology for the incorporation of interim analyses in SW-CRTs was recently described \citep{grayling2017b}. One other possible explanation is that the studies have been false negatives. A high false negative rate could be associated with SW-CRTs having been under-powered. Methodology available to determine the sample size required by SW-CRTs is dependent upon the specification of the values of several nuisance parameters (e.g., the between cluster and residual variances). In practice, it may be difficult to provide accurate estimates for these factors, and their mis-specification may be leading to under-powered studies. Alternatively, if these parameters are being mis-specified such that SW-CRTs have been over-powered, there may have been more measurements taken than actually required, leading to unnecessary cost.

A common approach to addressing the specification of nuisance parameters in the trial design literature is the use of a sample size re-estimation (SSRE) procedure. Each such method has essentially the same intention: to alleviate the issue of pre-specifying nuisance parameters by allowing them to be re-estimated during the trial, and the required sample size adjusted \citep{proschan2009}. Broadly speaking they can be sub-categorised into blinded or unblinded techniques, with regulatory agencies preferring to maintain the blind when possible so as to not risk compromising the validity of a trial \citep{ich1998}. Blinded SSRE methodology is today available for a range of settings (e.g., \citet{friede2013}, \citet{golkowski2014}, and \citet{kunz2017}), with each such procedure typically conferring highly desirable trial operating characteristics.

However, whilst some results exist on SSRE in multi-centre \citep{jensen2010} and parallel group CRTs \citep{vanschie2014}, no work has established methodology for SSRE in SW-CRTs, with the increased complexity in the design of SW-CRTs necessitating a specialised approach. In this article, we address this by developing and exploring the performance of both blinded and unblinded SSRE procedures for cross-sectional SW-CRTs. In particular a commonly considered linear mixed model will be utilised for data analysis, blinded estimators of the key variance parameters are developed. The performance of a SSRE procedure based on these blinded estimators is then compared to an unblinded approach, as a function of their various control parameters, and the parameters of the underlying model. We then conclude with a discussion of possible extensions to our approach, as well as logistical factors that must be considered when incorporating SSRE in to SW-CRTs.

\section{Methods}

\subsection{Notation, hypotheses and analysis}

We consider a scenario in which a cross-sectional SW-CRT is to be carried out in $C$ clusters over $T$ time periods, with $n$ individuals recruited per cluster per time period. That is, we assume data will be accrued on new patients in each cluster in each time period. We do not restrict our attention to `balanced' SW-CRTs however; clusters need not start in the control condition, conclude in the experimental condition, and there does not need to be an equal number of clusters switching to the experimental intervention in each time period.

We assume that the accumulated data will be normally distributed, and the following linear mixed model will be utilised for data analysis, as proposed by \citet{hussey2007}
\begin{equation}\label{eq2}
y_{ijk} = \mu + \pi_j + \tau X_{ij} + c_i + \epsilon_{ijk}.
\end{equation}
Here
\begin{itemize}
	\item $y_{ijk}$ is the response of the $k$th individual ($k=1,\dots,n$), in the $i$th cluster ($i=1,\dots,C$), in the $j$th time period ($j=1,\dots,T$);
	\item $\mu$ is an intercept term;
	\item $\pi_j$ is a fixed effect for the $j$th time period (with $\pi_1=0$ for identifiability);
	\item $\tau$ is a fixed treatment effect for the experimental intervention relative to the control;
	\item $X_{ij}$ is the binary treatment indicator for the $i$th cluster and $j$th time period. That is, $X_{ij}=1$ if cluster $i$ receives the intervention in time period $j$. We denote by $X$ the matrix formed from the $X_{ij}$, and by $X^{(t)}$ the first $t$ columns of $X$;
	\item $c_{i} \sim N(0,\sigma_c^2)$ is a random effect for cluster $i$;
	\item $\epsilon_{ijk} \sim N(0,\sigma_e^2)$ is the individual-level error.
\end{itemize}
We denote the vector of fixed effects by $\beta=(\mu,\pi_2,\dots,\pi_T,\tau)^T$. Moreover, we indicate the design matrix linking $\beta$ to the vector of responses $y_{\mathscr{T},n}$, from the set of time periods $\mathscr{T}$, given an allocation matrix $X$, and a per cluster per period sample size of $n$, by $D_{\mathscr{T},n}$. We similarly denote the covariance matrix of $y_{\mathscr{T},n}$, given $\sigma_c^2$ and $\sigma_e^2$, by $\text{Cov}(y_{\mathscr{T},n},y_{\mathscr{T},n} \mid\sigma_c^2,\sigma_e^2)=\Sigma_{\mathscr{T},n,\sigma_c^2,\sigma_e^2}$. As noted in \citet{hussey2007}, $\Sigma_{\mathscr{T},n,\sigma_c^2,\sigma_e^2}$ is an $|\mathscr{T}|n\times|\mathscr{T}|n$ block diagonal matrix.

We perform a one-sided hypothesis test for $\tau$
\[ H_0 : \tau \le 0, \qquad H_1 : \tau > 0, \]
and assume that it is desired to have a type-I error rate of $\alpha$ when $\tau=0$, and to have power to reject $H_0$ of $1-\beta$ when $\tau=\delta$, for some specified $\delta>0$. Note that SSRE procedures for two-sided hypotheses are also easily achievable.

Finally, we assume that hypothesised values for the variance parameters $\sigma_c^2$ and $\sigma_e^2$ have been provided, which we denote by $\tilde{\sigma}_c^2$ and $\tilde{\sigma}_e^2$. Alternatively, a value for one of these parameters, and a value for the intra-cluster correlation (ICC) $\rho$,  $\tilde{\rho} = \tilde{\sigma}_c^2/(\tilde{\sigma}_c^2 + \tilde{\sigma}_e^2)$, could be specified, such that $\tilde{\sigma}_c^2$ and $\tilde{\sigma}_e^2$ can still be determined. Given these values, we assume a sample size calculation has been performed (using the methods to be described shortly) and values for $X$ and $n$ (and thus also $C$ and $T$ since $\text{dim}⁡(X)=C\times T$) have subsequently been specified. For reasons to be elucidated below we refer to this $n$ as $n_\text{init}$.

With the above, a conventional SW-CRT can be conducted as follows. We recruit $n_\text{init}$ individuals per cluster per time period, with the experimental intervention allocated according to the matrix $X$. On completion, we use restricted error maximum likelihood (REML) estimation to acquire an estimate of $\hat{\tau}$, denoted $\hat{\tau}$, and a value for $\hat{I}=\{\text{Var}(\hat{\tau})\}^{-1}$. Next, we compute the test statistic $T=\hat{\tau}\hat{I}^{1/2}$, and reject $H_0$ if $T>e$, where e is the solution to
\begin{align*}
\alpha &= \int_e^{\infty}\varphi\{x,0,1,\nu\}\mathrm{d}x,\\
\nu &= n_\text{init}CT-C-T.
\end{align*}

Here, $\varphi\{x,\mu,\Lambda,\nu\}$ is the probability density function of a $t$-distribution with mean $\mu$, covariance $\Lambda$, and degrees of freedom $\nu$, evaluated at $x$. Moreover, specifically we take $\nu$ to be the degrees of freedom in a corresponding balanced multi-level ANOVA design. Later, we will discuss the implications of this and other possible ways to prescribe $\nu$.

We next detail how the above can be extended to allow SSRE to be incorporated in to the design.

\subsection{Sample size re-estimation procedures}\label{ssre}

A single SSRE interim analysis is included in the SW-CRT design after a designated time period $t\in\{1,\dots,T-1\}$. Specifically, we assume that the trial is conducted as per matrix $X$ and the value $n_\text{init}$ for time periods $1,\dots,t$. After this, we compute estimates for the variance parameters, $\hat{\sigma}_c^2$ and $\hat{\sigma}_e^2$, based upon the accumulated data. Below, we detail how exactly this is achieved in the blinded and unblinded procedures. Here, we discuss how these estimates are then used. 

Explicitly, we search numerically as follows to determine the required per cluster per period sample size for the remainder of the trial, $n_\text{reest}$, to convey the desired power if $\sigma_c^2=\hat{\sigma}_c^2$ and $\sigma_e^2=\hat{\sigma}_e^2$.  Thus, the number of clusters remains fixed throughout the trial; it is the per cluster per period sample size that is adjusted. We consider possible alternatives to this in the discussion.

Firstly, suppose $n_\text{reest}$ has been chosen, then time periods $t+1,\dots,T$ of the trial are conducted using the matrix $X$ for treatment allocation, and recruiting $n_\text{reest}$ individuals per cluster per period. At the end of the trial the linear mixed model (\ref{eq2}) with REML estimation are utilised to acquire $\hat{\tau}$ and $\hat{I}$ as above. The test statistic $T=\hat{\tau}\hat{I}^{1/2}$ is again determined, and $H_0$ rejected if $T>e$, but $e$ is now the solution to
\begin{align*}
\alpha &= \int_e^{\infty}\varphi\{x,0,1,\nu_{n_\text{reest}}\}\mathrm{d}x,\\
\nu_{n_\text{reest}} &= n_\text{init}Ct + n_\text{reest}C(T-t)-C-T.
\end{align*}
Here, $\nu_{n_\text{reest}}$ is the degrees of freedom in a balanced multi-level ANOVA design if a sample size of $n_\text{init}$ is used per cluster per period in time periods $1,\dots,t$, and a sample size of $n_\text{reest}$ is used per cluster per period in time periods $t+1,\dots,T$.

The power to reject $H_0$ when $\tau=\delta$, for a particular $n_\text{reest}$, can thus be estimated at the interim as
\[ \mathbb{P}(\text{Reject } H_0 \mid n_\text{reest}) = \int_e^{\infty}\varphi\{x,\delta I^{1/2},1,\nu_{n_\text{reest}}\}\mathrm{d}x,\]
where $I$ is given by the inverse of element $[T+1,T+1]$ of the following matrix
\begin{align*}
& \left(D_{\{1,\dots,t\},n_\text{init}}^T \Sigma_{\{1,\dots,t\},n_\text{init},\hat{\sigma}_c^2,\hat{\sigma}_e^2}^{-1} D_{\{1,\dots,t\},n_\text{init}} +\right. \\ & \qquad \qquad \left.D_{\{t+1,\dots,T\},n_\text{reest}}^T \Sigma_{\{t+1,\dots,T\},n_\text{reest},\hat{\sigma}_c^2,\hat{\sigma}_e^2}^{-1} D_{\{t+1,\dots,T\},n_\text{reest}}\right)^{-1}.
\end{align*}
This matrix arises as the theoretical covariance matrix of the maximum likelihood estimator of $\beta$ when a sample size of $n_\text{init}$ is used per cluster per period in time periods $1,\dots,t$, and a sample size of $n_\text{reest}$ is used per cluster per period in time periods $t+1,\dots,T$. 

Therefore, we can compute the required value for $n_\text{reest}$ by searching for the minimal integer solution to the following equation
\[ \mathbb{P}(\text{Reject } H_0 \mid n_\text{reest})\ge 1-\beta.\]
In fact, to make our SSRE procedures more applicable in practice, and to guard against unrealistically large values for $n_\text{reest}$, we carry out the remaining periods of the trial recruiting $n_\text{final}$ individuals per cluster per time period, where
\[ n_\text{final}=\begin{cases}
n_\text{min} &: n_\text{reest} < n_\text{min},\\
n_\text{reest} &: n_\text{min} \le n_\text{reest} \le n_\text{max},\\
n_\text{max} &: n_\text{max} < n_\text{reest}.
\end{cases} \]
Here, $n_\text{min}$ and $n_\text{max}$ are designated values for the minimal and maximal allowed number of patients per cluster per period following the re-estimation. These could be chosen for example based upon the practical attainable values of $n$ for a particular trial.

Finally, following determination of $n_\text{final}$, the remainder of the trial and ensuant analysis is conducted as described above, to determine whether to reject $H_0$.

Note that the sample size required by a classical fixed sample SW-CRT design, given an allocation matrix $X$, can be determined using the above by treating $n_\text{init}$ as a variable rather than a fixed parameter, and searching for the minimal $n_\text{init}$ such that $\mathbb{P}(\text{Reject } H_0 \mid 0)\ge 1-\beta$ when $t=T$. Alternatively, $n_\text{init}$ could be specified and the matrix $X$ determined for the desired power.

All that remains to be elucidated in the above procedure is the means of determining the estimates $\hat{\sigma}_c^2$ and $\hat{\sigma}_e^2$. As discussed, we describe both blinded and unblinded approaches to their specification.

The unblinded procedure is as follows. After time period $t$, we fit the following model to the accumulated data using REML estimation
\[ y_{ijk}=\begin{cases}
\mu + \pi_j + \tau X_{ij} + c_i + \epsilon_{ijk} &: \text{sum}(X^{(t)})>0 \text{ and } t>1,\\
\mu + \pi_j + c_i + \epsilon_{ijk} &: \text{sum}(X^{(t)})=0 \text{ and } t>1,\\
\mu + \tau X_{ij} + c_i + \epsilon_{ijk} &: \text{sum}(X^{(t)})>0 \text{ and } t=1,\\
\mu + c_i + \epsilon_{ijk} &: \text{sum}(X^{(t)})=0 \text{ and } t=1.
\end{cases} \]
Here, $\text{sum}(X^{(t)})>0$ is included as a qualifier to indicate the term $X_{ij}\tau$ should appear in our model as at least one cluster has been administered the experimental intervention in some time period. Similarly, $t>1$ indicates period effects should be accounted for in the model. From the REML estimator, we attain our values for $\hat{\sigma}_c^2$ and $\hat{\sigma}_e^2$ immediately, and use them in the above algorithm to determine $n_\text{final}$.

For the blinded procedure, we define
\begin{align*}
S_1^2 &= \sum_{i=1}^C \sum_{j=1}^t \sum_{k=1}^{n_\text{init}} (Y_{ijk} - \bar{Y}_{...}^{(t)})^2,\\
S_{Ct}^2 &= \sum_{i=1}^C \sum_{j=1}^t \sum_{k=1}^{n_\text{init}} (Y_{ijk} - \bar{Y}_{ij.})^2,
\end{align*}
where
\begin{align*}
\bar{Y}_{...}^{(t)} &= \frac{1}{n_\text{init}Ct}\sum_{i=1}^C \sum_{j=1}^t \sum_{k=1}^{n_\text{init}} Y_{ijk},\\
\bar{Y}_{ij.} &= \frac{1}{n_\text{init}}\sum_{k=1}^{n_\text{init}} Y_{ijk}.
\end{align*}
Then, in the Appendix we derive that in the absence of period effects
\begin{align*}
\mathbb{E}(S_1^2) &= \sigma_e^2 + \frac{n_\text{init}Ct}{n_\text{init}Ct-1}\left(1-\frac{1}{C}\right)\sigma_c^2 + \frac{n_\text{init}\tau^2}{n_\text{init}Ct-1}\text{sum}(X^{(t)}) \\
& \qquad - \frac{n_\text{init}^2\tau^2}{n_\text{init}Ct(n_\text{init}Ct-1)}\text{sum}(X^{(t)})^2,\\
\mathbb{E}(S_{Ct}^2) &= \sigma_e^2.
\end{align*}
Given a particular choice for $\tau$ in the above, which we shall denote $\tau_*$, these equations are used to estimate $\sigma_c^2$ and $\sigma_e^2$ as follows
\begin{itemize}
	\item Compute $S_1^2$ and $S_{Ct}^2$ using the formulae above and the accrued data.\\
	\item Define $f(S_1^2,\sigma_e^2,X^{(t)},n,\tau)$ as
	\begin{align*}
	f(S_1^2,\sigma_e^2,X^{(t)},n,\tau) &= \frac{nCt-1}{nCt}\frac{C}{C-1}\left\{S_1^2-\sigma_e^2-\frac{n\tau_*^2}{nCt-1}\text{sum}(X^{(t)})\right.\\ & \qquad \qquad \qquad \qquad \qquad  \left.+\frac{n^2\tau_*^2}{nCt(nCt-1)}\text{sum}(X^{(t)})^2 \right\}.
	\end{align*}
	\item Set $\hat{\sigma}_e^2=S_{Ct}^2$ and $\hat{\sigma}_c^2=\max\{f(S_1^2,\hat{\sigma}_e^2,X^{(t)},n_\text{init},\tau_*),0\}$.
\end{itemize}

We then utilise the algorithm from earlier for determining the value of $n_\text{final}$.

Note that therefore, in the absence of period effects, if $\tau_*=\tau$, the above are unbiased estimators for the variance parameters.

For further clarity, the full unblinded and blinded SSRE procedures are detailed algorithmically in the Appendix.

\subsection{Simulation study}\label{sim}

With the above considerations, a SSRE trial design scenario is fully specified given $\mathscr{D}$, where
\[ \mathscr{D}=\{X,t,\sigma_c^2,\sigma_e^2,\tilde{\sigma}_c^2,\tilde{\sigma}_e^2,\alpha,\beta,\delta,\mu,\boldsymbol{\pi},\tau,n_\text{min},n_\text{max},B\}\cup \mathbb{I}_{\{B=1\}}\{\tau_*\}.\]
Here, $\boldsymbol{\pi}=(\pi_2,\dots,\pi_T)^T$ is the vector of period effects, and $B$ is a binary indicator variable that takes the value 1 if blinded SSRE is utilised, and the value 0 if unblinded SSRE is utilised. Finally, $\mathbb{I}_A$ is the indicator function on event A.

Given $\mathscr{D}$ we can simulate a SW-CRT utilising this SSRE procedure by generating random multivariate normal observations. From this, the empirical rejection rate (ERR) of a particular scenario can be estimated by performing a large number of replicates simulations. To this end, define $R_s(\mathscr{D})$ to be 1 if the result of replicate $s$ of a trial simulated according to scenario $\mathscr{D}$ is to reject $H_0$, and 0 otherwise. For any number of replicates $r$, the ERR for scenario $\mathscr{D}$ is
\[ ERR(\mathscr{D})=\frac{1}{r} \sum_{s=1}^r R_s(\mathscr{D}).\]
In this article, $r=10^5$ for all considered scenarios.

In what follows, we consider the ERR in a wide variety of scenarios. However, many of the parameters in $\mathscr{D}$ remain fixed. In particular, they are set based on two motivating trial design scenarios.

Firstly, \citet{bashour2013} conducted a SW-CRT to assess the effect of training doctors in communication skills on women's satisfaction with doctor-woman relationship during labour and delivery. The trial utilised a balanced complete block SW-CRT design, enrolling four hospitals, and gathering data over five time periods. The final analysis estimated the between cluster and residual variances to be $\sigma_c^2=0.02$ and $\sigma_e^2=0.51$, respectively. For these variance parameters, the utilised design would have required 70 patients per cluster per time period for the trials desired type-I and type-II error rates of 0.05 and 0.1 respectively, when powering for a clinically relevant difference of 0.2, using the methods above. Thus in Trial Design Setting (TDS) 1 we fix $\sigma_c^2=0.02$, $\sigma_e^2=0.51$, $\alpha=0.05$, $\beta=0.1$, and $\delta=0.2$. Moreover, $C=4$, $T=5$, and $X$ is such that a single cluster switches to the experimental intervention in time periods two through five. Furthermore, we assume for this example that trialists would not want the per cluster per period sample size to drop post re-estimation, and thus set $n_\text{min}=n_\text{init}$. Finally, we always take $n_\text{max}=200$. This value was chosen as a fair compromise of an attainable $n$ (noting that the trial was easily able to recruit 100 patients per cluster per time period), and a value large enough such that for most considered values of $\tilde{\sigma}_c^2$, $\hat{\sigma}_e^2$ and $t$, the trial should be able to meet the power requirement.

The parameters of TDS2 are based upon the typical characteristics of SW-CRTs according to a recent review \citep{grayling2017}. Precisely, we take, as in \citet{grayling2017b}, $\sigma_c^2=1/9$, $\sigma_e^2=1$, $\alpha=0.05$, $\beta=0.2$ and $\delta=0.24$, to consider a more modest value for the ICC of $\rho=0.1$. In this case, to differ from TDS1, we do allow the per cluster per time period sample size to drop following re-estimation, setting $n_\text{min}=0.5n_\text{init}$ and $n_\text{max}=20$. Finally, we set $C=20$, $T=9$, and $X$ such that three clusters switch to the experimental intervention in time periods two through five, and two clusters in time periods six through nine.

Additionally, for simplicity, in both TDSs we take $\mu=0$ and $\tau_*=0$. We therefore consider the effect of different choices for $t$, $\tilde{\sigma}_c^2$, $\tilde{\sigma}_e^2$, $\boldsymbol{\pi}$, $\tau$, and $B$. Moreover, we primarily focus upon the following three combinations for the assumed variance parameters $\tilde{\sigma}_c^2$ and $\tilde{\sigma}_e^2$

\begin{itemize}
	\item Scenario 1: When the variance components are under-specified by 50\%; $\tilde{\sigma}_c^2=0.5\sigma_c^2$, $\tilde{\sigma}_e^2=0.5\sigma_e^2$.\\
	\item Scenario 2: When the variance components are correctly specified; $\tilde{\sigma}_c^2=\sigma_c^2$, $\tilde{\sigma}_e^2=\sigma_e^2$.\\
	\item Scenario 3: When the variance components are over-specified by 50\%; $\tilde{\sigma}_c^2=1.5\sigma_c^2$, $\tilde{\sigma}_e^2=1.5\sigma_e^2$.\\
\end{itemize}

Software to perform the above simulations is available from https://sites.google.com/site/jmswason/supplementary-material.

\section{Results}

\subsection{Performance for varying $\tilde{\sigma}_c^2$ and $\tilde{\sigma}_e^2$}

To begin, we consider how the SSRE procedures perform as $\sigma_c^2$ and $\sigma_e^2$ are mis-specified to varying degrees. For simplicity, we set $\boldsymbol{\pi}=\boldsymbol{0}$, and take $t=3$ and $t=5$ for TDS1 and TDS2 respectively. Precisely, we explore $(\tilde{\sigma}_c^2,\tilde{\sigma}_e^2) \in \{0.5\sigma_c^2,\sigma_c^2,1.5\sigma_c^2\}\times\{0.5\sigma_e^2,\sigma_e^2,1.5\sigma_e^2\}$, with $\tau=0$; the empirical type-I error rate (ETI), or $\tau=\delta$;  the empirical power (EP). Table 1 depicts the results, displaying the empirical rejection rates of the re-estimation procedures and the corresponding fixed sample SW-CRT design.

\renewcommand{\arraystretch}{1.2}
\begin{table}[htb]
	\begin{center}
		\caption{The empirical type-I error rate and power of the blinded and unblinded re-estimation procedures, along with the corresponding fixed sample SW-CRT design are shown. Results are given for trial design settings 1 and 2, for a selection of possible values for the assumed variance parameters, when $t=3$ and $t=5$ in trial design settings 1 and 2 respectively.}
		\begin{tabular}{ccccccccc}
			\hline
			& & \multicolumn{3}{c}{Empirical type-I error rate} && \multicolumn{3}{c}{Empirical power} \\
			\cline{3-5}\cline{7-9}
			$\tilde{\sigma}_c^2$ & $\tilde{\sigma}_e^2$ & Blinded & Unblinded & Fixed && Blinded & Unblinded & Fixed\\
			\hline
			\multicolumn{9}{c}{Trial Design Setting 1}\\
			\hline
			$0.5\sigma_c^2$ & $0.5\sigma_e^2$ & 0.0575 & 0.0572 & 0.0623 && 0.8527 & 0.8711 & 0.6905 \\
			$0.5\sigma_c^2$ & $\sigma_e^2$ & 0.0564 & 0.0570 & 0.0595 && 0.8993 & 0.9005 & 0.8930 \\
			$0.5\sigma_c^2$ & $1.5\sigma_e^2$ & 0.0544 & 0.0556 & 0.0577 && 0.9718 & 0.9714 & 0.9689 \\
			$\sigma_c^2$ & $0.5\sigma_e^2$ & 0.0564 & 0.0572 & 0.0622 && 0.8536 & 0.8712 & 0.7030 \\
			$\sigma_c^2$ & $\sigma_e^2$ & 0.0553 & 0.0565 & 0.0596 && 0.9055 & 0.9061 & 0.9024 \\
			$\sigma_c^2$ & $1.5\sigma_e^2$ & 0.0536 & 0.0529 & 0.0572 && 0.9760 & 0.9758 & 0.9748 \\
			$1.5\sigma_c^2$ & $0.5\sigma_e^2$ & 0.0577 & 0.0588 & 0.0642 && 0.8523 & 0.8719 & 0.7130 \\
			$1.5\sigma_c^2$ & $\sigma_e^2$ & 0.0572 & 0.0563 & 0.0579 && 0.9082 & 0.9105 & 0.9060 \\
			$1.5\sigma_c^2$ & $1.5\sigma_e^2$ & 0.0519 & 0.0531 & 0.0563 && 0.9780 & 0.9772 & 0.9738 \\
			\hline
			\multicolumn{9}{c}{Trial Design Setting 2}\\
			\hline
			$0.5\sigma_c^2$ & $0.5\sigma_e^2$ & 0.0505 & 0.0509 & 0.0526 && 0.7419 & 0.8039 & 0.6250 \\
			$0.5\sigma_c^2$ & $\sigma_e^2$ & 0.0499 & 0.0485 & 0.0523 && 0.7762 & 0.8108 & 0.8119 \\
			$0.5\sigma_c^2$ & $1.5\sigma_e^2$ & 0.0497 & 0.0495 & 0.0512 && 0.8529 & 0.8551 & 0.9076 \\
			$\sigma_c^2$ & $0.5\sigma_e^2$ & 0.0507 & 0.0493 & 0.0526 && 0.7415 & 0.8008 & 0.6221 \\
			$\sigma_c^2$ & $\sigma_e^2$ & 0.0500 & 0.0501 & 0.0512 && 0.7773 & 0.8069 & 0.8094 \\
			$\sigma_c^2$ & $1.5\sigma_e^2$ & 0.0491 & 0.0486 & 0.0507 && 0.8854 & 0.8846 & 0.9293 \\
			$1.5\sigma_c^2$ & $0.5\sigma_e^2$ & 0.0498 & 0.0492 & 0.0530 && 0.7390 & 0.8009 & 0.6247 \\
			$1.5\sigma_c^2$ & $\sigma_e^2$ & 0.0495 & 0.0489 & 0.0512 && 0.7761 & 0.8084 & 0.8118 \\
			$1.5\sigma_c^2$ & $1.5\sigma_e^2$ & 0.0486 & 0.0484 & 0.0518 && 0.8858 & 0.8851 & 0.9308 \\
			\hline
		\end{tabular}
	\end{center}
\end{table}

In general, assuming larger values for the variance parameters leads to an increased EP and a decreased ETI, as would be expected.

In TDS1, for certain values of the assumed variance parameters there is large inflation of the ETI above the nominal level. This is an issue common to both the SSRE procedures and the fixed design, with the maximal inflation observed for $\hat{\sigma}_c^2=1.5\sigma_c^2$, $\hat{\sigma}_e^2=0.5\sigma_e^2$ in the fixed design where the ETI is 0.0642. Finally, in this setting, the blinded procedure most often attains the smallest ETI, but this comes at a cost to its EP relative to the unblinded approach.

In contrast, for TDS2 the SSRE procedures routinely control the ETI to approximately the desired level, with the blinded and unblinded methods displaying similar ETIs. Though it is smaller than for TDS1, the fixed design still displays some inflation to the ETI. The unblinded procedure again typically has higher EP than the blinded approach. However, the difference between the two procedures EP is now more pronounced. In fact, in most instances, the blinded procedure does not attain the desired power. In this case, because the SSRE procedures are allowed to lower the per cluster per period sample size at the interim analysis, for some assumed values of the variance parameters they have an EP smaller than the fixed design.

Overall, it is clear that for many assumed values of the two variance components, the SSRE procedures have a far higher EP than the corresponding fixed SW-CRT design, with comparable if not preferable ETIs. For example, when $\tilde{\sigma}_c^2=0.5\sigma_c^2$ and $\tilde{\sigma}_e^2=0.5\sigma_e^2$ in TDS2, the blinded procedure has an EP of 0.8039, whilst the corresponding conventional SW-CRT design has an EP of only 0.6250; an increase of 29\%. 

\subsection{Performance for varying $t$}

Next, we assess the impact upon the ETI and EP of the choice of the SSRE point $t$. As above, we set $\boldsymbol{\pi}=\boldsymbol{0}$. However, we now focus on the three assumed variance scenarios given in Section~\ref{sim}.

Table 2 displays the ETI and EP of the blinded and unblinded SSRE procedures when $\tau=0$ and $\tau=\delta$ respectively, for $t\in\{2,3,4\}$ in TDS1 and $t\in\{3,5,7\}$ in TDS2. There is no observable trend to the ETI as $t$ is increased in either TDS. For most of the considered assumed values for the variance parameters, the ETI is comparable for each $t$.

For TDS1, $t=3$ leads to the largest EP in all instances. In TDS2, the allowance to lower the per cluster per period sample size results in each of the considered values for $t$ attaining the maximal power for some assumed variance combination.

In some cases, it is clear that placing the re-estimation point later in to the trial can cause a substantial loss of power. For example, when $\tilde{\sigma}_c^2=0.5\sigma_c^2$ and $\tilde{\sigma}_e^2=0.5\sigma_e^2$ in TDS1, the unblinded procedure has power close to the nominal level of 0.8711 when $t=3$, but this drops to 0.7690 for $t=4$.

\begin{table}[htb]
	\begin{center}
		\caption{The empirical type-I error rate and empirical power of the blinded and unblinded re-estimation procedures are shown. Precisely, results are given for trial design settings 1 and 2, for a selection of possible values for the assumed variance parameters, as a function of the re-estimation time point $t$.}
		\begin{tabular}{cccccccccc}
			\hline
			& & & \multicolumn{3}{c}{Empirical type-I error rate} && \multicolumn{3}{c}{Empirical power} \\
			\cline{4-6}\cline{8-10}
			\multicolumn{10}{c}{Trial Design Setting 1}\\
			\hline
			Procedure & $\tilde{\sigma}_c^2$ & $\tilde{\sigma}_e^2$ & $t=2$ & $t=3$ & $t=4$ && $t=2$ & $t=3$ & $t=4$\\
			\hline
			Blinded & $0.5\sigma_c^2$ & $0.5\sigma_e^2$ & 0.0597 & 0.0575 & 0.0507 && 0.8454 & 0.8527 & 0.7682 \\
			Blinded & $\sigma_c^2$ & $\sigma_e^2$ & 0.0566 & 0.0553 & 0.0582 && 0.9049 & 0.9055 & 0.9050 \\
			Blinded & $1.5\sigma_c^2$ & $1.5\sigma_e^2$ & 0.0558 & 0.0519 & 0.0545 && 0.9762 & 0.9780 & 0.9765 \\
			Unblinded & $0.5\sigma_c^2$ & $0.5\sigma_e^2$ & 0.0620 & 0.0572 & 0.0499 && 0.8582 & 0.8711 & 0.7690 \\
			Unblinded & $\sigma_c^2$ & $\sigma_e^2$ & 0.0567 & 0.0565 & 0.0578 && 0.9076 & 0.9061 & 0.9057 \\
			Unblinded & $1.5\sigma_c^2$ & $1.5\sigma_e^2$ & 0.5238 & 0.0531 & 0.0536 && 0.9760 & 0.9772 & 0.9757 \\
			\hline
			\multicolumn{10}{c}{Trial Design Setting 2}\\
			\hline
			Procedure & $\tilde{\sigma}_c^2$ & $\tilde{\sigma}_e^2$ & $t=3$ & $t=5$ & $t=7$ && $t=3$ & $t=5$ & $t=7$\\
			\hline
			Blinded & $0.5\sigma_c^2$ & $0.5\sigma_e^2$ & 0.0514 & 0.0505 & 0.0479 && 0.7323 & 0.7419 & 0.7251 \\
			Blinded & $\sigma_c^2$ & $\sigma_e^2$ & 0.0495 & 0.0500 & 0.0491 && 0.7637 & 0.7773 & 0.7959 \\
			Blinded & $1.5\sigma_c^2$ & $1.5\sigma_e^2$ & 0.0480 & 0.0486 & 0.0490 && 0.8391 & 0.8858 & 0.9151 \\
			Unblinded & $0.5\sigma_c^2$ & $0.5\sigma_e^2$ & 0.0515 & 0.0509 & 0.0477 && 0.8019 & 0.8039 & 0.7475 \\
			Unblinded & $\sigma_c^2$ & $\sigma_e^2$ & 0.0502 & 0.0501 & 0.0503 && 0.8108 & 0.8069 & 0.8067 \\
			Unblinded & $1.5\sigma_c^2$ & $1.5\sigma_e^2$ & 0.0481 & 0.0484 & 0.0502 && 0.8393 & 0.8851 & 0.9145 \\
			\hline
		\end{tabular}
	\end{center}
\end{table}

\subsection{Performance for varying $\boldsymbol{\pi}$}

In the above, we have considered only $\boldsymbol{\pi}=\boldsymbol{0}$. This is a useful scenario to explore, since often period effects will be expected to be, and ultimately will be, small. However, the estimators used in our blinded SSRE procedure are only unbiased in the absence of period and treatment effects, whilst our unblinded SSRE procedure is only asymptotically invariant to the value of these effects. It is thus important to assess the effect of non-zero period effects on the SSRE procedures. Here, we explore this for $\tau=0$.

Precisely, we consider the ETI of our SSRE procedures when the value of $\pi_j$ ($j=1,\dots,T$) in each replicate scenario is drawn randomly as $\pi_j \sim N(0,\sigma_\pi^2)$. Note that these are unscaled $\pi_j$, we still force $\pi_1=0$ for identifiability purposes when fitting the linear mixed model~(\ref{eq2}). We then conduct these simulations for several values of $\sigma_\pi^2$, to determine how the strength of the period effects affects the ETI. Moreover, we achieve this for the three assumed variance scenarios given in Section~\ref{sim}, taking $t=3$ in TDS1 and $t=5$ in TDS2.

Figures 1 and 2 display the results of these simulations. It is evident that, allowing for Monte Carlo error, the value of $\sigma_\pi^2$ appears to have little impact upon the ETI as $\sigma_\pi^2$ is increased for any of the considered scenarios. 

Moreover, performance is clearly comparable between the blinded and unblinded procedures, if not arguably better in the blinded approach.

\begin{figure}[htb]
	\begin{center}\label{pi1}
		\includegraphics[width=14cm]{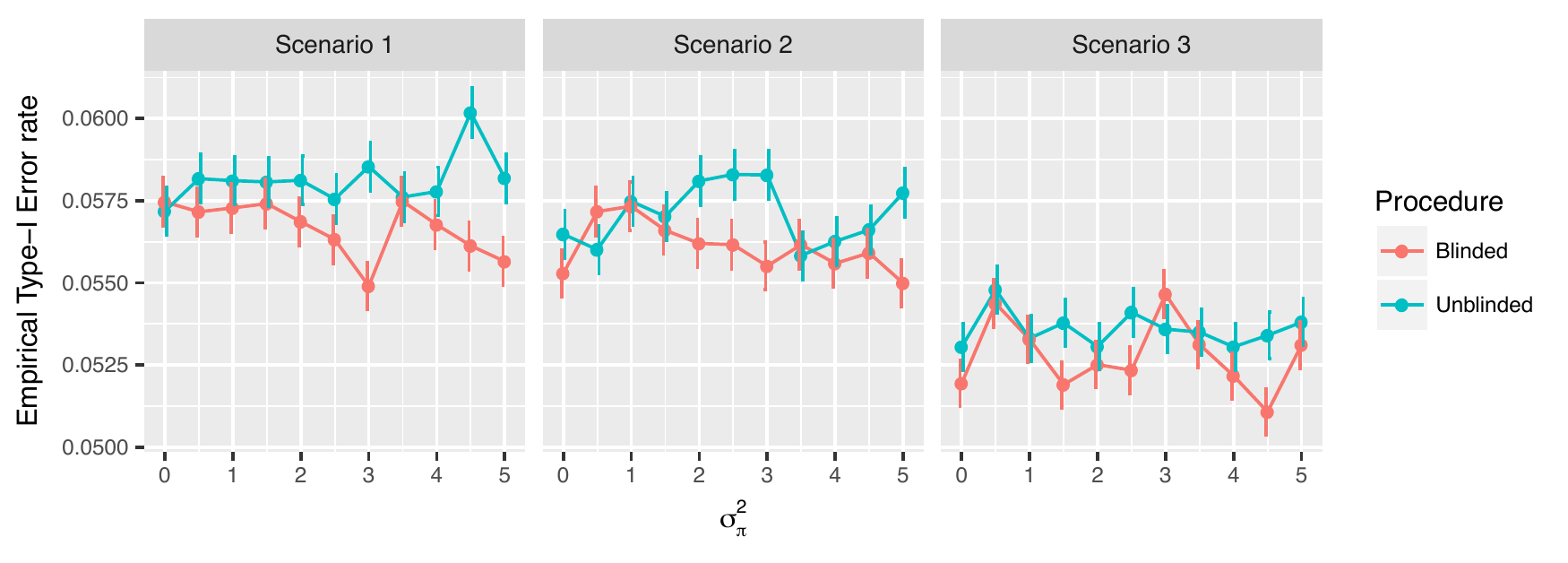}
		\caption{The empirical type-I error rate of the blinded and unblinded sample size re-estimation procedures is shown for the three considered variance parameter scenarios in TDS1 as a function of the variance of the period effects $\sigma_\pi^2$. Error bars indicate the Monte Carlo error.}
	\end{center}
\end{figure}

\begin{figure}[htb]
	\begin{center}\label{pi2}
		\includegraphics[width=14cm]{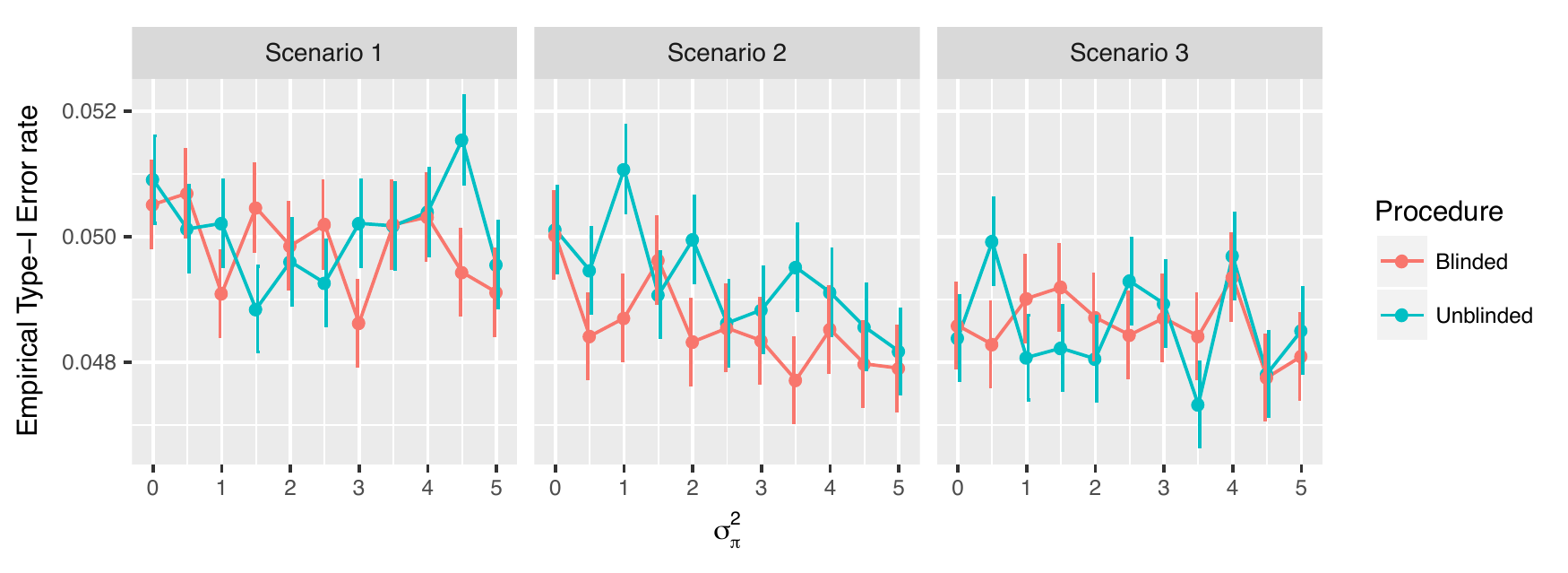}
		\caption{The empirical type-I error rate of the blinded and unblinded sample size re-estimation procedures is shown for the three considered variance parameter scenarios in TDS2 as a function of the variance of the period effects $\sigma_\pi^2$. Error bars indicate the Monte Carlo error.}
	\end{center}
\end{figure}

\section{Discussion}\label{disc}

In this article, we have presented blinded and unblinded SSRE procedures for cross-sectional SW-CRTs. These methods should assist with scenarios in which there is difficultly in determining a trial's required sample size because of the need to specify values for several nuisance parameters. We were able to demonstrate that, at least for the considered scenarios, the SSRE procedures could increase power substantially over a conventional SW-CRT design when the variance parameters were under-estimated.

Unfortunately, in TDS1 there were instances of substantial inflation to the ETI rate using our SSRE procedures. This was not surprising given the extremely low number of clusters in this scenario, with past research highlighting issues in such a setting \citep{taljaard2016,grayling2017b}. Additionally, it follows results observed for parallel-group CRTs \citep{vanschie2014}. To address this, one could use the Kenward-Roger approach to the specification of the degrees of freedom in the final analysis \citep{kenward1997}. Alternatively, an alpha-level adjustment procedure, as considered for example by \citet{golkowski2014} could be utilised. With either of these methods, one may anticipate that the the type-I error rate could be controlled more closely to the nominal level, but the design still attain a higher power than the corresponding fixed sample approach.

Unfortunately, in TDS2 the blinded SSRE routinely did not display an EP above the nominal level. To combat this, one could employ a sample size inflation factor, as proposed by \citet{zucker1999}. This has been demonstrated to be highly effective in a range of trial design setting (e.g., \citet{friede2013,golkowski2014}). Using it, it could be possible for the preferable blinded SSRE procedure to provide the desired power.

We observed that the ETI and EP were similar for several choices of $t$, particularly in TDS1, but the EP was sometimes substantially lower if the re-estimation point was late in the trial. This should not be surprising, since a larger value of $t$ implies less time to re-adjust for any mis-specifications in the variance parameters. Of course, a smaller value for $t$ implies less data has been accumulated, and so we may expect on average less accurate estimates for $\sigma_c^2$ and $\sigma_e^2$. One may therefore suggest an intermediate option, such as $t=3$ in TDS1, to be preferable.

Though we anticipated that the unblinded procedure would have more desirable properties in the case of non-zero period effects, we actually found that the performance of the two types of SSRE procedure were similar. Consequently, it may be possible even in the case of strong period effects for the preferable blinded SSRE procedure to be utilised. Of course, researchers should always extensively examine the operating characteristics of any SSRE procedure in a range of scenarios before utilisation to verify this to be the case.

As well as determining the influence of non-zero period effects, this should also include assessing a designs sensitivity to the choice of $n_\text{min}$ and $n_\text{max}$. In particular, whilst it may be preferable to have $n_\text{min}<n_\text{init}$, this could have negative consequences upon the EP. This has been discussed previously for conventional parallel arm individually randomised trials (see, for example, \citet{bowden2014}). It was evident here in TDS2, where for example the blinded SSRE procedure had an EP below the desired level when $\sigma_e^2$ was specified correctly, but the fixed design did not. This does however confer an advantage that when $\tilde{\sigma}_e^2=1.5\sigma_e^2$ the SSRE procedures were able to reduce the power to closer to the nominal level compared to the fixed design. Likewise, increasing the value of $n_\text{max}$ may seem beneficial, but one then needs both to be able to find more patients to recruit in the later periods, and also to be able to logistically handle a larger sample size. This may be a problem particularly for scenarios where the SW-CRT design is being utilised because of resource constraints.

There are several practical factors that must be considered before SSRE is incorporated in to a SW-CRT design. Primarily, our methodology is dependent upon data from all clusters being available for analysis immediately following period $t$. The efficiency of the procedures would suffer if this were not the case. Therefore, it would be important for measures to be put in place for efficient data collection, storage, and analysis. In addition, there may be some instances where SSRE is not realistic. For example, if the intervention was a planned roll-out that is part of a larger programme implementation. A trialist must consider their scenario carefully before utilising SSRE.

Several possible extensions to our procedures are possible. Firstly, we here only addressed cross-sectional SW-CRT designs analysed with the Hussey and Hughes model. Though the majority of SW-CRT research has been set in this domain, it would be beneficial to also establish methods to incorporate SSRE in to cohort designed SW-CRT, different endpoints of interest, or indeed different analysis models. Whilst it would be relatively simple to explore the performance of an unblinded procedure in these settings, methodology for blinded re-estimation would be more complex. Similar statements also hold for allowing variable cluster sizes, and also incorporating the interim estimated value for $\tau$ in to the re-estimation procedure.

Additionally, we considered here a scenario in which the number of clusters remained fixed throughout the trial; adjusting only the per cluster per period sample size following the re-estimation point. One could also explore the performance of a procedure that increases the value of $C$ following re-estimation, creating an incomplete-block SW-CRT. For scenarios in which patients are hard to come by, but clusters are not, this would be a useful extension.

It is worth noting that our procedures are actually applicable to any cross-sectional CRT design to be analysed with the Hussey and Hughes model. This means, for example, that it would allow also the incorporation of SSRE in to a cluster randomised crossover trial, which is being increasingly acknowledged in the trials community as a useful design \citep{arnup2014}.

Regardless of the practical considerations discussed above, and the possible future avenues of extension to our methods, it is clear that the ability to include a SSRE point in to SW-CRT designs is a useful addition to the methodologists toolbox.

\section{Acknowledgement}
	This work was supported by the Wellcome Trust [grant number 099770/Z/12/Z to M.J.G.]; the Medical Research Council [grant number MC\_UP\_1302/2 to A.P.M.]; and the National Institute for Health Research Cambridge Biomedical Research Centre [grant number MC\_UP\_1302/4 to J.M.S.W.].
\vspace*{1pc}

\noindent {\bf{Conflict of Interest}}

\noindent {\it{The authors have declared no conflict of interest.}}

\section*{Appendix}

\subsection*{A.1.\enspace Blinded estimators}

In this section, we derive the expected value of the blinded estimators given in Section~\ref{ssre}.

To begin, we observe that for equation~(\ref{eq2}), for $i_1,i_2\in\{1,\dots,C\}$, $j_1,j_2\in\{1,\dots,T\}$, and $k_1,k_2\in\{1,\dots,n\}$
\[\text{Cov}(Y_{i_1j_1k_1},Y_{i_2j_2k_2}) = \delta_{i_1i_2}\delta_{j_1j_2}\delta_{k_1k_2}\sigma_e^2 + \delta_{i_1i_2}\sigma_c^2,\]
as is stated in \citet{hussey2007}.
Moreover, using the standard ``." notation to indicate when a variable has been summed over, and denoting $\mathbb{S}_a = \{(x,y)\in(1,\dots,a)\times(1,\dots,a) : x\neq y\}$
\begin{align*}
\text{Cov}\left(\bar{Y}_{ij.},\bar{Y}_{ij.}\right) &= \text{Var}\left(\bar{Y}_{ij.}\right),\\
&= \text{Var}\left(\frac{1}{n}\sum_{k=1}^{n} Y_{ijk}\right),\\
&= \frac{1}{n^2}\text{Var}\left(\sum_{k=1}^{n} Y_{ijk}\right),\\
&= \frac{1}{n^2}\left[\sum_{k=1}^{n}\text{Var}(Y_{ijk}) + \sum_{(k_1,k_2)\in\mathbb{S}_n}\text{Cov}\left(Y_{ijk_1},Y_{ijk_2}\right) \right],\\
&= \frac{1}{n^2}\left[(\sigma_e^2 + \sigma_c^2) + \sum_{(k_1,k_2)\in\mathbb{S}_a}\sigma_c^2 \right],\\
&= \frac{1}{n^2}\left[n(\sigma_e^2 + \sigma_c^2) + n(n-1)\sigma_c^2 \right],\\
&= \frac{1}{n}\left[\sigma_e^2 + n\sigma_c^2\right].
\end{align*}
Additionally, taking
\[ N_t = \sum_{i=1}^{C}\sum_{j=1}^{t}n=nCt,\]
with $N=N_T$, and recalling $X^{(t)}$ is the matrix formed by restricting $X$ to its first $t$ columns, if
\[ \bar{Y}_{...}^{(t)} = \frac{1}{N_t} \sum_{i=1}^C \sum_{j=1}^{t} \sum_{k=1}^{n}Y_{ijk}, \]
we have that
\begin{align*}
\text{cov}\left(\bar{Y}_{...}^{(t)},\bar{Y}_{...}^{(t)}\right) &= \text{var}\left(\bar{Y}_{...}^{(t)}\right),\\
&= \text{var}\left(\frac{1}{N_t}\sum_{i=1}^{C}\sum_{j=1}^{t}\sum_{k=1}^{n} Y_{ijk}\right),\\
&= \frac{1}{N_t^2}\text{var}\left(\sum_{i=1}^{C}\sum_{j=1}^{t}\sum_{k=1}^{n} Y_{ijk}\right),\\
\begin{split}
&= \frac{1}{N_t^2}\left[\sum_{i=1}^{C}\sum_{j=1}^{t}\sum_{k=1}^{n}\text{var}(Y_{ijk})\right. \\
& \qquad \qquad \left. + \sum_{(i_1,i_2)\in\mathbb{S}_C}\sum_{(j_1,j_2)\in\mathbb{S}_t}\sum_{(k_1,k_2)\in\mathbb{S}_n}\text{cov}\left(Y_{i_1j_1k_1},Y_{i_2j_2k_2}\right) \right],
\end{split}\\
&= \frac{1}{N_t^2}\left[N_t(\sigma_e^2 + \sigma_c^2) + nCt(nt-1)\sigma_c^2 \right],\\
&= \frac{1}{N_t}\left(\sigma_e^2 + \frac{N_t}{C}\sigma_c^2\right).
\end{align*}
Now, after time period $t$, two sensible variances can be computed; the variance of all response values gathered thus far (the one sample variance, $S_1^2$), and the variance of the response values from each cluster in each time period thus far from their corresponding mean values (the $Ct$ sample variance, $S_{Ct}^2$). Explicitly, we have
\begin{align*}
(N_t-1)S_1^2 &= \sum_{i=1}^C\sum_{j=1}^t\sum_{k=1}^n\left(Y_{ijk} - \bar{Y}_{...}^{(t)} \right)^2,\nonumber\\
&= \left(\sum_{i=1}^C\sum_{j=1}^t\sum_{k=1}^nY_{ijk}^2\right) - N_t\bar{Y}_{...}^{(t)2},\\
(N_t-Ct)S_{Ct}^2 &= \sum_{i=1}^C\sum_{j=1}^t\sum_{k=1}^n\left(Y_{ijk} - \bar{Y}_{ij.} \right)^2,\nonumber\\
&= \sum_{i=1}^C\sum_{j=1}^t\left[\sum_{k=1}^n Y_{ijk}^2 - n\bar{Y}_{ij.}^2 \right].
\end{align*}
Without loss of generality, we can assume that $\mu=0$. By definition, we know that
\[ \mathbb{E}\left[(N-Ct)S_{Ct}^2\right] = (N-Ct)\sigma_e^2.\]
Furthermore, exploiting the fact that $X_{ij}^2=X_{ij}$
\begin{align*}
\mathbb{E}\left[(N_t-1)S_1^2\right] &= \left[\sum_{i=1}^C\sum_{j=1}^t\sum_{k=1}^n\mathbb{E}(Y_{ijk}^2)\right] - N_t\mathbb{E}(\bar{Y}_{...}^{(t)2}),\\
\begin{split}
&= \left\{\sum_{i=1}^C\sum_{j=1}^t\sum_{k=1}^n\left[\text{var}(Y_{ijk})+\mathbb{E}(Y_{ijk})^2\right]\right\} \\
& \qquad - N_t\left[\text{var}(\bar{Y}_{...}^{(t)}) + \mathbb{E}(\bar{Y}_{...}^{(t)})^2\right],\\
\end{split}\displaybreak[0]\\
\begin{split}
&= \left\{\sum_{i=1}^C\sum_{j=1}^t\sum_{k=1}^n\left[(\sigma_e^2+\sigma_c^2)+(\pi_j + X_{ij}\tau)^2\right]\right\}\\
& \qquad - N_t\left\{\frac{1}{N_t}\left(\sigma_e^2 + \frac{N_t}{C}\sigma_c^2\right) + \left[\frac{1}{N_t}\sum_{i=1}^C\sum_{j=1}^t\sum_{k=1}^n(\pi_j+X_{ij}\tau)\right]^2\right\},
\end{split}\displaybreak[0]\\
\begin{split}
&= N_t(\sigma_e^2 + \sigma_c^2) + \sum_{i=1}^C\sum_{j=1}^t\sum_{k=1}^n (\pi_j + X_{ij}\tau)^2 - \left(\sigma_e^2 + \frac{N_t}{C}\sigma_c^2\right)\\
& \qquad - \frac{1}{N_t}\left[\sum_{i=1}^C\sum_{j=1}^t\sum_{k=1}^n(\pi_j+X_{ij}\tau)\right]^2,\\
\end{split}\displaybreak[0]\\
\begin{split}
&= (N_t-1)\sigma_e^2 + N_t\left(1 - \frac{1}{C}\right)\sigma_c^2 + \sum_{i=1}^C\sum_{j=1}^t\sum_{k=1}^n \pi_j^2 \\
& \qquad + \sum_{i=1}^C\sum_{j=1}^t\sum_{k=1}^n (X_{ij}\tau)^2 + 2\sum_{i=1}^C\sum_{j=1}^t\sum_{k=1}^n \pi_jX_{ij}\tau \\
& \qquad - \frac{1}{N_t}\left(\sum_{i=1}^C\sum_{j=1}^t\sum_{k=1}^n\pi_j\right)^2 - \frac{1}{N_t}\left(\sum_{i=1}^C\sum_{j=1}^t\sum_{k=1}^nX_{ij}\tau\right)^2 \\
& \qquad - \frac{2}{N_t}\left(\sum_{i=1}^C\sum_{j=1}^t\sum_{k=1}^n\pi_j\right)\left(\sum_{i=1}^C\sum_{j=1}^t\sum_{k=1}^nX_{ij}\tau\right),
\end{split}\displaybreak[0]\\
\begin{split}
&= (N_t-1)\sigma_e^2 + N_t\left(1 - \frac{1}{C}\right)\sigma_c^2 + nC\sum_{j=1}^t\pi_j^2 \\
& \qquad + n\tau^2\sum_{i=1}^C\sum_{j=1}^tX_{ij}^2 + 2n\tau\sum_{i=1}^C\sum_{j=1}^t\pi_jX_{ij} \\
& \qquad - \frac{n^2C^2}{N_t}\left(\sum_{j=1}^t\pi_j\right)^2 - \frac{n^2\tau^2}{N_t}\left(\sum_{i=1}^C\sum_{j=1}^tX_{ij}\right)^2 \\
& \qquad - \frac{2n^2C\tau}{N_t}\left(\sum_{j=1}^t\pi_j\right)\left(\sum_{i=1}^C\sum_{j=1}^tX_{ij}\right),
\end{split}\displaybreak[0]\\
\begin{split}
&= (N_t-1)\sigma_e^2 + N_t\left(1 - \frac{1}{C}\right)\sigma_c^2 + nC(\boldsymbol{\pi}^{(t)}\cdot\boldsymbol{\pi}^{(t)}) \\
& \qquad + n\tau^2\text{sum}(X^{(t)}) + 2n\tau \text{sum}(X^{(t)}\boldsymbol{\pi}^{(t)}) \\
& \qquad - \frac{n^2C^2}{N_t}\text{sum}(\boldsymbol{\pi}^{(t)})^2 - \frac{n^2\tau^2}{N_t}\text{sum}(X^{(t)})^2 \\
& \qquad - \frac{2n^2C\tau}{N_t}\text{sum}(\boldsymbol{\pi}^{(t)})\text{sum}(X^{(t)}),
\end{split}
\end{align*}
where $\boldsymbol{\pi}^{(t)}=(\pi_1,\dots,\pi_t)^T$, and $\text{sum}(A)$ for a matrix $A$ indicates the sum of all of its elements. Therefore, in the absence of period effects ($\boldsymbol{\pi}^{(t)}=\boldsymbol{0}$), we have
\begin{align*}
\mathbb{E}\left(S_1^2\right) &= \sigma_e^2 + \frac{N_t}{N_t-1}\left(1 - \frac{1}{C}\right)\sigma_c^2+\frac{n\tau^2}{N_t-1}\text{sum}(X^{(t)}) - \frac{n^2\tau^2}{N_t(N_t-1)}\text{sum}(X^{(t)})^2,\\
\mathbb{E}(S_{Ct}^2) &= \sigma_e^2,
\end{align*}
as given in the main part of the paper.

\subsection{Sample size re-estimation procedures: algorithm}

Here, we provide a complete point-by-point algorithm for how the blinded and unblinded re-estimation procedures should be conducted.

Firstly, our blinded SSRE procedure is as follows
\begin{enumerate}
	\item Specify values for $X$, $\alpha$, $\beta$, $\delta$, $\tilde{\sigma}_c^2$, $\tilde{\sigma}_e^2$, $t$, $\tau_*$, $n_{\text{min}}$ and $n_{\text{max}}$.
	\item Perform an initial sample size determination, to acquire $n_{\text{init}}$, assuming $\sigma_e^2=\tilde{\sigma}_e^2$ and $\sigma_c^2=\tilde{\sigma}_c^2$.
	\item Conduct the trial up to the end of time period $t$, recruiting $n_{\text{init}}$ individuals per cluster per period.
	\item Compute $S_1^2$ and $S_{Ct}^2$.
	\item Set $\hat{\sigma}_e^2 = S_{Ct}^2$ and $\hat{\sigma}_c^2=\max\{f(S_1^2,\hat{\sigma}_e^2,X^{(t)},n_\text{init},\tau_*),0\}$.
	\item Compute the exact required per cluster per period sample size, $n_{\text{reest}}$, for the rest of the trial to imply the desired operating characteristics assuming $\sigma_c^2=\hat{\sigma}_c^2$ and $\sigma_e^2=\hat{\sigma}_e^2$. Then, set $n_{\text{final}}$ as follows
	\[ n_\text{final}=\begin{cases}
	n_\text{min} &: n_\text{reest} < n_\text{min},\\
	n_\text{reest} &: n_\text{min} \le n_\text{reest} \le n_\text{max},\\
	n_\text{max} &: n_\text{max} < n_\text{reest}.
	\end{cases} \]
	\item Conduct periods $t+1,\dots,T$ of the trial, recruiting $n_{\text{final}}$ patients per cluster per period.
	\item Perform a final unblinded analysis on all accumulated data using equation~(\ref{eq2}) to determine efficacy.
\end{enumerate}

Finally, our unblinded SSRE procedure is as follows

\begin{enumerate}
	\item Specify values for $X$, $\alpha$, $\beta$, $\delta$, $\tilde{\sigma}_c^2$, $\tilde{\sigma}_e^2$, $t$, $\tau_*$, $n_{\text{min}}$ and $n_{\text{max}}$.
	\item Perform an initial sample size determination, to acquire $n_{\text{init}}$, assuming $\sigma_e^2=\tilde{\sigma}_e^2$ and $\sigma_c^2=\tilde{\sigma}_c^2$.
	\item Conduct the trial up to the end of time period $t$, recruiting $n_{\text{init}}$ individuals per cluster per period.
	\item Fit the following model to all accumulated data using REML estimation
	\[ y_{ijk} = \begin{cases}
	\mu + c_i + \pi_j + X_{ij}\tau + \epsilon_{ijk}& : \text{if } \text{sum}(X^{(t)})>0\text{ and }t>1,\\
	\mu + c_i + \pi_j + \epsilon_{ijk}& : \text{if } \text{sum}(X^{(t)})=0\text{ and }t>1,\\
	\mu + c_i + X_{ij}\tau + \epsilon_{ijk} & : \text{if } \text{sum}(X^{(t)})>0\text{ and }t=1,\\
	\mu + c_i + \epsilon_{ijk} & : \text{if } \text{sum}(X^{(t)})=0\text{ and }t=1.
	\end{cases}. \]
	\item From the fitted model obtain the estimates $\hat{\sigma}_c^2$ and $\hat{\sigma}_e^2$.
	\item Compute the exact required per cluster per period sample size, $n_{\text{reest}}$, for the rest of the trial to imply the desired operating characteristics assuming $\sigma_c^2=\hat{\sigma}_c^2$ and $\sigma_e^2=\hat{\sigma}_e^2$. Then, set $n_{\text{final}}$ as follows
	\[ n_\text{final}=\begin{cases}
	n_\text{min} &: n_\text{reest} < n_\text{min},\\
	n_\text{reest} &: n_\text{min} \le n_\text{reest} \le n_\text{max},\\
	n_\text{max} &: n_\text{max} < n_\text{reest}.
	\end{cases} \]
	\item Conduct periods $t+1,\dots,T$ of the trial, recruiting $n_{\text{final}}$ patients per cluster per period.
	\item Perform a final unblinded analysis on all accumulated data using equation~(\ref{eq2}) to determine efficacy.
\end{enumerate}

\end{document}